\documentclass{emulateapj}

\usepackage{amsmath}
\usepackage{mathrsfs}
\newcommand{\bs}[1]{\boldsymbol{#1}}

\makeatletter

\begin{document}

\shorttitle{Non-linear Rees-Sciama: Analytical vs Simulation}
\title{Cosmic Microwave Background--Weak Lensing Correlation:
Analytical and Numerical Study of Non-linearity 
and Implications for Dark Energy}

\author{Atsushi J. Nishizawa\altaffilmark{1}, 
  Eiichiro Komatsu\altaffilmark{2},
  Naoki Yoshida\altaffilmark{1},
  Ryuichi Takahashi\altaffilmark{1}
  Naoshi Sugiyama\altaffilmark{1,3}}

\altaffiltext{1}{Department of Physics, Nagoya University, Furocho, Nagoya, 
  Aichi 464-8602, Japan}
\altaffiltext{2}{Department of Astronomy, University of Texas at Austin, 
  Austin, TX 78712 USA}
\altaffiltext{3}{Institute for Physics and Mathematics of the Universe, 
  University of Tokyo, 5-1-5 Kashiwa-no-Ha, Kashiwa City, 
  Chiba 277-8582, Japan}

\shortauthors{Nishizawa et al.}

\email{atsushi@a.phys.nagoya-u.ac.jp}

%%%%%%%%%%%%%%%%%%%%%%%%%%%%%%%%%%%%%%%%%%%%%%%%%%%%%%%%%%%%%%%%%%%%%%%%%%%%%%%
%%%%%%%%%%%%%%%%%%%%%%%%%%%%%%%%%%%%%%%%%%%%%%%%%%%%%%%%%%%%%%%%%%%%%%%%%%%%%%%
\begin{abstract} 
%%%%%%%%%%%%%%%%%%%%%%%%%%%%%%%%%%%%%%%%%%%%%%%%%%%%%%%%%%%%%%%%%%%%%%%%%%%%%%%
%%%%%%%%%%%%%%%%%%%%%%%%%%%%%%%%%%%%%%%%%%%%%%%%%%%%%%%%%%%%%%%%%%%%%%%%%%%%%%%
Non-linear evolution of density fluctuations 
yields secondary anisotropies in the cosmic microwave
background (CMB), which are correlated with the same density fluctuations
that can be measured by weak 
lensing (WL) surveys. We study the CMB-WL correlation
using analytical models as well as $N$-body simulations. 
We show that an analytical model based upon the time derivative of 
non-linear matter power spectrum agrees with simulations.
All-sky cosmic-variance limited CMB and WL surveys allow us to
measure the correlation from non-linearity with high
significance ($50\sigma$) for $l_{\rm max}=10^4$, whereas the
forthcoming missions such as Planck and LSST are expected to yield only
marginal detections.
The CMB-WL correlation is sensitive to the time derivative of structure 
growth.
We study how this property may be used to 
constrain the nature of dark energy. 
While  
the expected constraints are not very strong, they may provide a cross
check of results from other observations.
\end{abstract}


%%%%%%%%%%%%%%%%%%%%%%%%%%%%%%%%%%%%%%%%%%%%%%%%%%%%%%%%%%%%%%%%%%%%%%%%%%%%%%%
%%%%%%%%%%%%%%%%%%%%%%%%%%%%%%%%%%%%%%%%%%%%%%%%%%%%%%%%%%%%%%%%%%%%%%%%%%%%%%%
\section{Introduction}
\label{sec:intro}
%%%%%%%%%%%%%%%%%%%%%%%%%%%%%%%%%%%%%%%%%%%%%%%%%%%%%%%%%%%%%%%%%%%%%%%%%%%%%%%
%%%%%%%%%%%%%%%%%%%%%%%%%%%%%%%%%%%%%%%%%%%%%%%%%%%%%%%%%%%%%%%%%%%%%%%%%%%%%%%
Secondary temperature anisotropies of the cosmic microwave background (CMB)
provide invaluable information on the structure formation in the universe.
The sources of anisotropies include the Sunyaev-Zel'dovich (SZ)
effects by galaxy clusters as well as by reionization, the Rees-Sciama
(RS) effect, and CMB lensing. 

The forthcoming Planck satellite and
ground-based observations such as Atacama Cosmology Telescope 
\citep[ACT;][]{Kosowsky:2003} and South-Pole Telescope \citep[SPT;][]{SPT:2004}
are designed to measure temperature fluctuations at arc-minute
scales, and thus are expected to detect some of the secondary
anisotropies. 

The Rees-Sciama effect is, in principle, a unique probe of the 
{\it time-variation} of gravitational potential, as it is given by
\citep{SachsWolfe:1967, ReesSciama:1968}
\begin{equation}
  \frac{\Delta T(\hat{\bs{n}})}{T} = 2\int_0^{r_*} dr
  \Phi'(\hat{\bs{n}}r,r),
  \label{eq:1}
\end{equation}
where $\Phi$ is the Newtonian potential, 
$\hat{\bs{n}}$ is a unit direction vector, $r$ is the
conformal {\it lookback} 
time, and $r_*$ is $r$ out to the photon decoupling epoch.
The prime denotes
$\partial/\partial r$, which is equal to $-\partial/\partial\eta$,
where $\eta$ is the conformal time.

In the linear regime, $\Phi'$ vanishes in a matter-dominated
universe; however, when either
curvature or dark energy dominates, $\Phi$ decays in time and thus
$|\Phi|' > 0$. 
In the {\it non-linear} regime, on the other hand, $\Phi$ {\it grows} in
time, $|\Phi|' < 0$. 
Therefore, one expects $\Phi\Phi'>0$
on large scales where density fluctuations are still linear,
and $\Phi\Phi'<0$ on small scales where fluctuations are non-linear.

The auto-correlation of the RS effect, both linear and non-linear, has
been studied \citep{Seljak:1996,Tuluie.etal:1996,Cooray:2002a}.
The cross-correlation between the {\it linear} 
RS effect and large-scale structure
traced by galaxies has also been studied
\citep[e.g.,][]{Crittenden:1995,Peiris:2000,Cooray:2002b,Afshordi:2004}, and
detected for various galaxy surveys
\citep[e.g.,][]{Boughn:2003,Nolta:2003,Afshordi.etal:2004}.

Much less attention has been given to the correlation of
the {\it non-linear} RS effect and large-scale structure.
As this effect changes the sign at the linear-to-non-linear 
transition scale, 
the signature of non-linearity is distinct.

The RS effect is not the only thing that is
correlated with the large-scale structure. The SZ effects, as well as
radio and infrared point sources, also trace the large-scale
structure. Fortunately, 
multi-frequency data enable us to separate these contributions that have
unique and specific frequency dependence. 

In this paper, we calculate the cross-correlation 
of the RS effect and large-scale structure traced by weak lensing (WL) surveys.
In particular, we simulate this directly using a $N$-body code for the
first time. 

We adopt the standard $\Lambda$CDM model with 
$(\Omega_{m0},\Omega_{\Lambda0}, \sigma_8, h, n_s)$
$=(0.26, 0.74, 0.76, 0.7, 1)$,
which is 
consistent with \citet{spergel.etal:2006}.

%%%%%%%%%%%%%%%%%%%%%%%%%%%%%%%%%%%%%%%%%%%%%%%%%%%%%%%%%%%%%%%%%%%%%%%%%%%%%%%
%%%%%%%%%%%%%%%%%%%%%%%%%%%%%%%%%%%%%%%%%%%%%%%%%%%%%%%%%%%%%%%%%%%%%%%%%%%%%%%
\section{Analytical Model}
\label{sec:cl}
%%%%%%%%%%%%%%%%%%%%%%%%%%%%%%%%%%%%%%%%%%%%%%%%%%%%%%%%%%%%%%%%%%%%%%%%%%%%%%%
%%%%%%%%%%%%%%%%%%%%%%%%%%%%%%%%%%%%%%%%%%%%%%%%%%%%%%%%%%%%%%%%%%%%%%%%%%%%%%%

The amplitude of distortion in galaxy images due to WL 
is given by the so-called
convergence, $\kappa$, which is proportional to the projected 
density field along the line of sight.
As the Newtonian potential, $\Phi$, is negatively proportional to density 
field via the Poisson equation, 
the sign of CMB-WL correlation is given by
$-\Phi\Phi'$; thus, negative on large scales and positive on small scales.

The cross-correlation of the RS effect and $\kappa$, $C_l$, is 
\begin{align}
    \label{eq:2}
    C_l
    = 
    -2l^2 \int_{0}^{r_*} dr P_{\Phi\Phi'}(l/r;r)
    \int_{z}^{z_*}dz_s~p(z_s) \frac{r_s-r}{r_s r^3},
\end{align}
where $P_{\Phi\Phi'}(k;r)$ is the power spectrum of $\Phi\Phi'$ at 
$r$, $z_s$ is the redshift of source galaxies, 
and $r_s\equiv r(z_s)$.
Here, $p(z)dz$ is the probability of finding galaxies between $z$ and $z+dz$. 
We use 
$p(z)=A z^2 \exp[-(z/z_0)^{\beta}]$, where 
$A$ is a normalization factor determined by $\int_0^\infty p(z)dz=1$
\citep{Efstathiou:1991}.
We consider two survey designs:
({\it Model 1}) Deep Survey, $(\beta,z_0)=(0.7, 0.5)$, 
whose $p(z)$ peaks at $z\sim 2.2$ with a broad distribution, 
and
({\it Model 2}) Shallow Survey, 
 $(\beta,z_0)=(2,0.9)$,
which peaks at $z\simeq 0.9$ with a narrower distribution.

We calculate $P_{\Phi\Phi'}$ from
\begin{equation}
    P_{\Phi\Phi'}(k,r)= \frac{9\Omega_{m0}^2 H_0^4}{4a^2k^4}
    \left[P_{\delta\delta'}(k,r)-{\mathscr{H}}P_{\delta\delta}(k,r)\right],
    \label{eq:3}
\end{equation}
where $\mathscr{H}=-d\ln a / dr$, and
$P_{\delta\delta}$ and $P_{\delta\delta'}$ are the power spectrum of 
density fluctuations, $\delta$, and the cross-correlation of $\delta$ and
$\delta'$, respectively.

How do we calculate $P_{\delta \delta'}$?
The continuity equation,  $\delta'=q$ where
 $q\equiv \nabla \cdot [\bs{v}(1+\delta)]$ is the momentum divergence, 
gives the {\it exact} relation, 
$P_{\delta \delta'}=P_{\delta q}$. Here,
$P_{\delta q}$ is the 
density-momentum divergence cross spectrum.

It might be tempting to use
$P_{\delta \delta'}=P_{\delta\delta}'/2$, as the following might
seem obvious:
$    \langle {\delta}_{\bs{k}}(r)
    {\delta}'_{\bs{p}}(r) \rangle
    =
    \frac{1}{2}\frac{\partial}{\partial r}
    \langle {\delta}_{\bs{k}}(r)
        {\delta}_{\bs{p}}(r)\rangle$.
However, this relation is exact only in the linear regime, for which
the ensemble average (taken over initial perturbations) and
$\partial/\partial r$ commute. 
As they do not generally commute in the non-linear regime, we
check whether this {\it ansatz} is a good
approximation in the non-linear regime.

We test this ansatz by 
comparing $P_{\delta q}$ and $P'_{\delta\delta}/2$.
We use the 3rd-order perturbation theory (PT),
as it allows us
to study this delicate issue analytically.

We expand $\delta_{\bs{k}}(r)$ in a series up to the 3rd
order in the initial linear perturbation, $\delta_1(\bs{k})$, as
${\delta}_{\bs{k}}(r) = \sum_{n=1}^{3} D^n(r) \delta_n(\bs{k})$.
We also expand the velocity divergence, $\theta\equiv  \nabla
\cdot\bs{v}$, in Fourier space up to the 3rd
order in the initial perturbation,
$\theta_1(\bs{k})$, as
${\theta}_{\bs{k}}(r) = \sum_{n=1}^{3} D'(r)D^{n-1}(r) \theta_n(\bs{k})$.
Here, $D$ is the linear growth factor, and $\delta_n$ and $\theta_n$ 
 are of order $\delta_1^n$ and $\theta_1^n$, respectively. 

With this expansion, we obtain
\begin{eqnarray}
\nonumber
    P_{\delta q}(k,r)&=&
\nonumber 
     D(r)D'(r)P^{11}_{\delta\delta} (k) \\
\nonumber
&+&
     D^3(r)D'(r)\left[P^{13}_{\delta\theta}(k)+
     P^{31}_{\delta\theta}(k)+ P^{22}_{\delta\theta}(k)\right]\\
\nonumber
&+&
     D^3(r)D'(r)\int\frac{d^3p}{(2\pi)^3}\frac{\bs{k}\cdot\bs{p}}{p^2}
     \left[B^{112}_{\delta\theta\delta}(\bs{k},\bs{p})\right.\\
& &\left.\qquad\qquad+
     B^{121}_{\delta\theta\delta}(\bs{k},\bs{p})+
     B^{211}_{\delta\theta\delta}(\bs{k},\bs{p})
     \right],
    \label{eq:4}
\end{eqnarray}
where 
$P_{\delta\delta}^{11}(k)$ is the initial linear spectrum, 
$P_{\delta\theta}^{ij}(k)$ is the cross-correlation of 
$\delta_i$ and $\theta_j$, and
$B^{ijl}_{\delta\theta\delta}(\bs{k},\bs{p})\equiv 
\langle \delta_i(\bs{k}-\bs{p})
\theta_j(\bs{p})\delta_l(\bs{k})\rangle$, whose
explicit forms can be calculated in a straightforward manner by following
\citet{Scoccimarro:2004}.

We then compare $P_{\delta q}$ with the
3rd order expression of $P_{\delta\delta}'/2$
\citep{Bernardeau:2001} 
\begin{eqnarray}
\nonumber
    \frac12P'_{\delta\delta} (k,r)
    &=& 
    D(r)D'(r)P^{11}_{\delta\delta}(k)\\
    \label{eq:5}
    &+&
    2D^3(r)D'(r)[P^{22}_{\delta\delta}(k) +2P^{13}_{\delta\delta}(k)].
\end{eqnarray}
While equation (\ref{eq:4}) and (\ref{eq:5}) are obviously identical to the
first-order, it is not immediately clear whether 
$P_{\delta q}=P_{\delta\delta}'/2$ holds in the non-linear regime. 
We find $P_{\delta q}\approx P_{\delta\delta}'/2$
to the accuracy better than
10\% for $k < 1 h/$Mpc, and 5\% for $k < 0.4 h/$Mpc, at $r=0$.

In the right panel of Figure \ref{fig:compareNL}, we compare
$C_l$ calculated from 
$P_{\delta q}$ (dashed) and $P_{\delta\delta}'/2$ (solid).
We find that, in the non-linear regime ($l\gtrsim 1000$)
where the correlation turns positive, $C_l$ from $P_{\delta\delta}'/2$ 
is smaller by $\sim 10\%$.
This result suggests that we can 
use $P_{\delta\delta'}=P_{\delta\delta}'/2$ to
compute $P_{\Phi\Phi'}$ from equation (\ref{eq:3})
reasonably accurately.

However, our argument is still limited to PT, which
breaks down for $\delta\gtrsim 1$. 
In the same panel of Figure \ref{fig:compareNL}, the dotted lines show
$C_l$ from $P_{\delta\delta}'$, where
$P_{\delta\delta}$ is the fully evolved non-linear power spectrum
\citep{Smith.etal:2003}.
The 3rd-order PT overestimates
non-linearity in $P_{\delta\delta}$
\citep{Bernardeau:2001,JeongKomatsu2006}, and thus PT
predicts that the sign of $C_l$ changes at smaller $l$. This result
shows limitations of PT. 

\section{Analytical Model vs $N$-body Simulation}

To test $P_{\delta \delta'}\approx P_{\delta\delta}'/2$ in the
non-linear regime, we compare the analytical model with 
$N$-body simulations. 

We employ $512^3$ dark matter particles in a
volume of $L_{\rm box}=250 h^{-1}$Mpc and $40 h^{-1}$Mpc on a side. 
The simulations are performed by the GADGET-2 code
\citep{Springel:2005}. The initial  
conditions are generated by a standard method using the Zel'dovich 
approximation. 
We dump outputs from $z = 0.01$ to $10$ uniformly sampled in $\log (1+z)$.
For each output redshift, we dump adjacent two outputs in order to calculate 
$\delta'$ and $\Phi'$.

\begin{figure*}
    \begin{center}
        \plotone{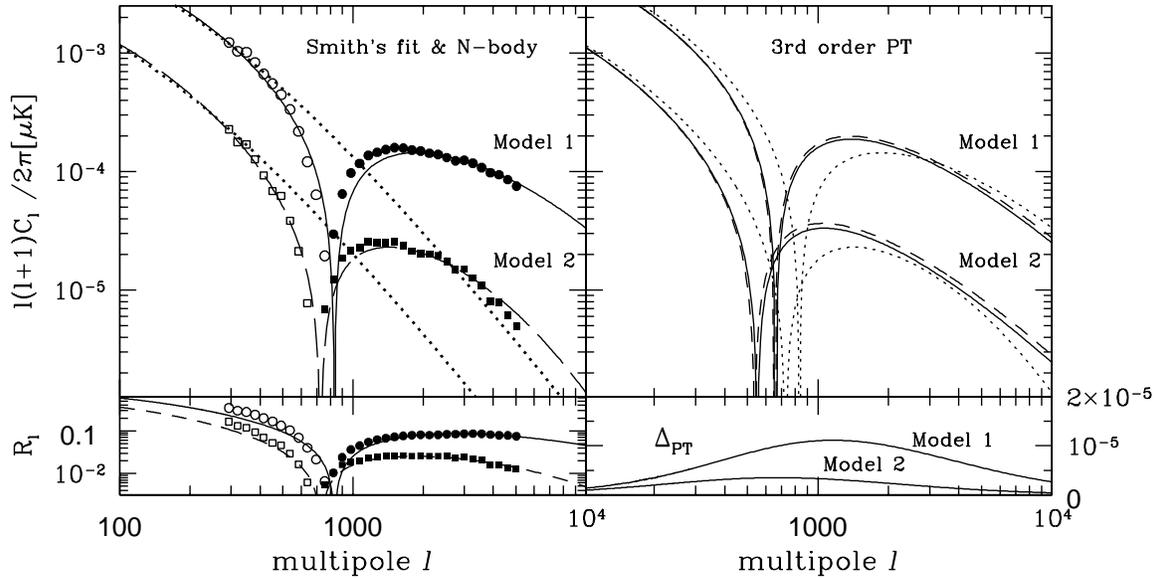}
        \caption{
          CMB-WL cross-correlation spectra, $C_l$, for two
          WL survey designs: Deep (Model 1) and Shallow (Model 2).  
({\it Top left}) 
          The symbols show $C_l$ from the $N$-body simulation.
          The open and filled symbols show $C_l<0$ and $C_l>0$, respectively.
          The solid and long-dashed lines show the fully non-linear
     model for Model 1 and 2, respectively, while the dotted lines show
     the linear theory predictions. Note that the linear theory
          predicts $C_l<0$ at all $l$. 
({\it Bottom left}) CMB-WL cross-correlation coefficients, $R_l$.
({\it Top right}) 3rd-order PT predictions. The solid and
          dashed lines show two ways of 
          calculating $C_l$, based upon Eq.~(\ref{eq:5}) and
          (\ref{eq:4}), respectively.          
          The fully non-linear
          calculations are also shown as the dotted lines (which are exactly
          the same as the solid and long-dashed lines on the top-left panel).
({\it Bottom right}) Difference between two ways of
          calculating $C_l$. The fractional differences
          are at most $\sim 10$\%.}
        \label{fig:compareNL}
    \end{center}
\end{figure*}

In the top left panel of Figure \ref{fig:compareNL},
we show the analytical model of $C_l$ with Smith et al.'s 
$P_{\delta\delta}$ (solid and long-dashed show Model 1 and 2, respectively) 
and the $N$-body results (open and
filled symbols show negative and positive values, respectively).
The agreement is good: the model describes 
the amplitude, shape, and 
cross-over at $l \sim 800$ of $C_l$ that are measured 
in the simulation. We therefore conclude that
 the ansatz, 
$P_{\delta\delta'}\approx P_{\delta\delta}'/2$,
is indeed accurate, up to $l=5000$ where we can trust resolution
of our $N$-body simulation.

How well are RS and WL correlated?
In the bottom left panel of Figure \ref{fig:compareNL}, 
we show the 2-d correlation coefficient, 
$R_l = C_l/\sqrt{C_l^{\kappa}C_l^{RS}}$.
We have used the same non-linear $P_{\delta\delta}$ to calculate
the power spectrum of convergence, $C_l^{\kappa}$, while we have used 
the halo model approach \citep{CooraySheth:2002}
to calculate that of RS, $C_l^{RS}$.
The RS and $\kappa$ are
strongly anti-correlated, $R_l \simeq -1$, in the linear regime.
(It is not exactly $-1$ because of the projection effect.)
The correlation weakens as non-linearity becomes
important: $R_l\simeq 0.05-0.1$ at $10^3\lesssim l\lesssim 10^4$ 
for Deep (Model 1), and  
$R_l\simeq 0.01-0.03$ for Shallow (Model 2).

The weak correlation of $C_l$ is due to the fact that
$\Phi$ and $\Phi'$ are not correlated very well in the non-linear
regime: the 3-d correlation coefficient, $R_{\rm 
3D}(k)=P_{\Phi\Phi'}(k)/\sqrt{P_{\Phi}(k)P_{\Phi'}(k)}$, reaches the
maximal value, $R_{\rm 3D}(0.5/{\rm Mpc})\simeq 0.2$, 
at $z=1$, where the RS effect becomes largest.
The 2-d correlation, $R_l$, is even weaker than $R_{\rm 3D}(k)$
because of the projection
effect and a mismatch between the redshift at which WL becomes
largest ($z\simeq 0.5$) and that for RS ($z\simeq 1$).

The weak correlation makes it challenging to
measure the CMB-WL 
correlation from non-linearity.

%%% fig:compareNL

%%%%%%%%%%%%%%%%%%%%%%%%%%%%%%%%%%%%%%%%%%%%%%%%%%%%%%%%%%%%%%%%%%%%%%%%%%%%%%%
%%%%%%%%%%%%%%%%%%%%%%%%%%%%%%%%%%%%%%%%%%%%%%%%%%%%%%%%%%%%%%%%%%%%%%%%%%%%%%%
\section{Detecting non-linear Rees--Sciama Effect}
\label{sec:detection}
%%%%%%%%%%%%%%%%%%%%%%%%%%%%%%%%%%%%%%%%%%%%%%%%%%%%%%%%%%%%%%%%%%%%%%%%%%%%%%%
%%%%%%%%%%%%%%%%%%%%%%%%%%%%%%%%%%%%%%%%%%%%%%%%%%%%%%%%%%%%%%%%%%%%%%%%%%%%%%%

Can we detect non-linearity, $C_l>0$?
The signal-to-noise ratio ($S/N$) of $C_l$ is given by
\begin{eqnarray}
    \left(\frac{S}{N}\right)^2_{l_{\rm max}}
    = f_\mathrm{sky}\sum_{l=2}^{l_{\rm max}}
\frac{2l+1}{1+\left(\tilde{C}_l^{\rm CMB}\tilde{C}_l^{\kappa}/C_l^2\right)},
    \label{eq:6}
\end{eqnarray}
where $f_{\rm sky}$ is the fraction of sky observed, and
$\tilde{C}^{CMB}_l$ and $\tilde{C}^\kappa_l$ are the 
total (signal plus noise) power spectra of CMB (including primary, CMB
lensing and RS) 
and WL, respectively. The noise spectra of CMB and WL surveys are
given, respectively, by \citep{Knox:1995,Schneider:2005}
\begin{eqnarray}
    N_l^{\rm CMB}
    &=& 
    \sigma_{\rm pix}^2\theta_{\rm fwhm}^2 
    \exp[l(l+1)\theta^2_{\rm fwhm}/8\ln 2], \\
    N_l^{\kappa}
    \label{eq:7} 
    &=&
    \sigma_{\gamma}^2/n_{\rm gal},
    \label{eq:8}
\end{eqnarray}
where $\sigma_{\rm pix}$ is the temperature noise per pixel, 
$\theta_{\rm fwhm}$ is the FWHM of a Gaussian beam,  
$n_{\rm gal}$ is the number density of galaxies 
observed in a WL survey, and $\sigma_{\gamma}$ is the noise in shear
measurements, including intrinsic ellipticities.

We forecast $S/N$ for the forthcoming surveys with the following parameters:
          ($f_{\rm sky}$, $\sigma_{\gamma}$, $n_g/{\rm arcmin}^2$)=
          (0.024, 0.3, 20) for Canada-France-Hawaii Telescope Legacy
          Survey (CFHTLS), (0.8, 0.1, 100) for 
Large Synoptic Survey Telescope (LSST), 
          ($f_{\rm sky}$, $\sigma_{\rm pix}$, 
          $\theta_{\rm FWHM}$) = (0.024, 4.4$\mu$K, $1'.7$) for ACT, 
          and (0.8, 1.7/2/4/3$\mu$K, $10'.7$/$8'$/$5'.5$) for Planck 
          with 3 channels.

\begin{figure}
    \begin{center}
        \plotone{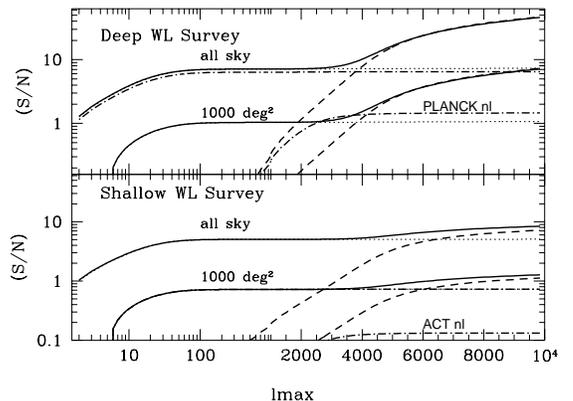}
        \caption{
          Signal-to-noise ratio ($S/N$) as a function of the
          maximum multipole, $l_{max}$. The dotted and dashed lines show
          the linear and non-linear contributions, respectively, while the
          solid lines show the total, for cosmic-variance limited
          CMB and WL surveys with all-sky (upper curves) and 1000-deg$^2$ sky
          coverage (lower curves).
({\it Top}) Deep WL Survey. The dashed-dotted lines show $S/N$
          expected from Planck's CMB data correlated with LSST's WL data.
          ``PLANCK nl'' shows the non-linear contribution only.
({\it Bottom}) Shallow WL Survey. Same as the top panel, but for
          ACT's CMB data correlated with CFHTLS's WL data.}
        \label{fig:sn}
    \end{center}
\end{figure}

In Figure \ref{fig:sn}, we show the predicted $S/N$ as a function of the
maximum multipole, $l_{\rm max}$, for cosmic-variance limited
CMB and WL (Deep and Shallow) surveys, as well as for the
forthcoming surveys: Planck correlated with LSST, and ACT correlated
with CFHTLS. 

We find that $S/N$ is totally dominated by the linear
contribution (dotted lines) 
at $l\lesssim 3000$, and then becomes dominated by
the non-linear contribution (dashed lines) at higher $l$.
All-sky CMB and WL surveys can yield
$S/N\sim 50$ (10) for Deep (Shallow) WL Survey, whereas 1000 deg$^2$
surveys can only yield $S/N\sim 7$ (1). 

Once noise of the forthcoming surveys is included, however, 
$S/N$ from the non-linear contribution becomes small
compared to the linear contribution. For Planck+LSST we find $S/N\sim
1.5$ for the non-linear, and 6 for the linear. For ACT+CFHTLS we find
$S/N\sim 0.1$ for the non-linear, and 0.7 for the linear. 
Therefore, we conclude that these forthcoming surveys are not expected
to yield significant detection of non-linear RS effect.

%%% fig:sn

%%%%%%%%%%%%%%%%%%%%%%%%%%%%%%%%%%%%%%%%%%%%%%%%%%%%%%%%%%%%%%%%%%%%%%%%%%%%%%%
%%%%%%%%%%%%%%%%%%%%%%%%%%%%%%%%%%%%%%%%%%%%%%%%%%%%%%%%%%%%%%%%%%%%%%%%%%%%%%%
\section{Sensitivity to dark energy parameters}
\label{sec:fim}
%%%%%%%%%%%%%%%%%%%%%%%%%%%%%%%%%%%%%%%%%%%%%%%%%%%%%%%%%%%%%%%%%%%%%%%%%%%%%%%
%%%%%%%%%%%%%%%%%%%%%%%%%%%%%%%%%%%%%%%%%%%%%%%%%%%%%%%%%%%%%%%%%%%%%%%%%%%%%%%

As the linear RS effect vanishes during the matter era, it is 
sensitive to dark energy (DE) 
\citep[e.g.,][]{Boughn:2003,Nolta:2003,Afshordi.etal:2004}.
The non-linear effect measures the structure growth, which is also sensitive 
to DE \citep{VerdeSpergel:2002}. The cross-over
at $l \sim 800$, at which the linear and non-linear contributions
cancel, is particularly a unique probe of DE.

\begin{figure}
    \begin{center}
        \plotone{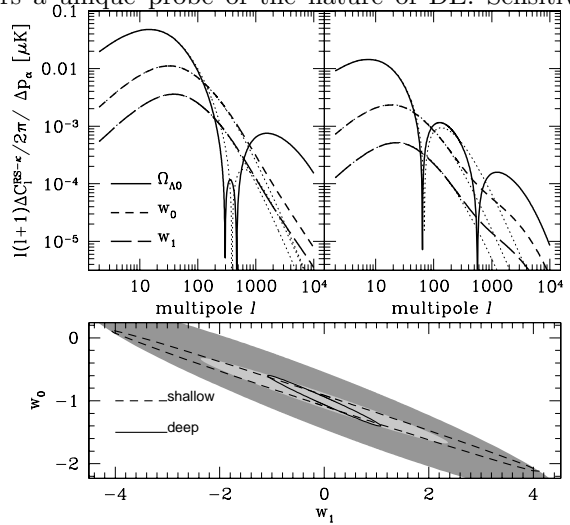}
        \caption{Sensitivity of the CMB-WL correlation to dark energy (DE).
({\it Top}) Derivatives of $C_l$ with respect to the DE parameters.
        The solid, short-dashed, and long-dashed curves show
        $\partial C_l/\partial\Omega_{\Lambda 0}$,  
        $\partial C_l/\partial w_0$, and $\partial
        C_l/\partial w_1$, respectively. The thin dotted lines show those from
        linear theory.
({\it Bottom}) Fisher matrix forecasts for all-sky cosmic-variance
        limited CMB and WL surveys. The light and dark areas
        show the expected 1-$\sigma$ 
        constraints from Deep and Shallow WL Surveys, respectively, with
        $\Omega_{\Lambda 0}$ marginalized.  
        The solid and dashed lines show the
        same things with $\Omega_{\Lambda 0}$ fixed.}
      \label{fig:clkt_cospar}
  \end{center}
\end{figure}

The top panels of Figure \ref{fig:clkt_cospar} show sensitivity of $C_l$
to DE parameters: $\partial C_l/\partial\Omega_{\Lambda 0}$,
$\partial C_l/\partial w_0$, and $\partial C_l/\partial w_1$.
We use a simple form of DE equation of
state, $w(a)=p_{\Lambda}(a)/\rho_{\Lambda}(a)$, given by
$w(a) = w_0 + w_1(1-a)$. The fiducial values are $w_0=-1$ and $w_1=0$.

We find  $\partial C_l/\partial w_0 <0$ and $\partial
C_l/\partial w_1 <0$ at all $l$. By increasing $w_0$ or $w_1$, one makes
$w(a)$ less negative which, in turn, makes DE more important at
earlier times. This does two things. On large scales where $C_l<0$, it
enhances the linear RS effect, making $C_l$ even more negative. On small
scales where $C_l>0$, it reduces the growth of non-linear 
structure, making $C_l$ less positive. 
In both cases we find negative derivatives.

Dependence on $\Omega_{\Lambda 0}$ is more complex.
While $\partial C_l/\partial \Omega_{\Lambda 0} <0$ at most $l$
can be explained by the same physics as above,
we  find  $\partial C_l/\partial \Omega_{\Lambda 0} >0$ 
in the intermediate $l$.
This is due to the assumption of a flat universe: 
a larger $\Omega_{\Lambda 0}$ results in a smaller $\Omega_{\rm m0}$, which
alters the shape of $P_{\delta\delta}(k)$
by shifting it to larger scales. (A smaller $\Omega_{\rm m0}$ delays 
matter-radiation equality.) This shift causes the intermediate $l$ to behave
differently. The behavior in the intermediate $l$ has little
to do with non-linearity, as we observe 
the same effect in the linear prediction.

Having evaluated the derivatives, we use 
the Fisher matrix analysis to calculate the expected constraints on
DE parameters, $p_\alpha=(\Omega_{\Lambda 0},~w_0,~w_1)$,
 from the CMB-WL correlation.
The Fisher matrix is given by
$    F_{\alpha \beta}
    = 
    \sum_{l=2}^{l_{\rm max}}
(\partial C_l/\partial p_{\alpha})
    {\rm Cov}^{-1}
(\partial C_l/\partial p_{\beta})$
where Cov is the covariance matrix (Eq.~\ref{eq:6})
and $l_{\rm max}=10^4$.

%%%% fig:clkt_cospar

The bottom panel of Figure \ref{fig:clkt_cospar} shows the expected joint
constraints on $w_0$ and $w_1$ from $C_l$ of all-sky cosmic-variance
limited CMB and WL surveys.
When $\Omega_{\Lambda 0}$ is not marginalized but fixed at the fiducial
value, we find $w_0 = -1 \pm 1.1~(0.4)$ and $w_1 = 0 \pm 4.1~(1.1)$ for
Shallow (Deep) WL Survey.
The constraints are not very strong, unfortunately; however, 
 the direction of degeneracy line, $w_1 \simeq 2(w_0+1)$, is about twice as 
steep as that of a joint analysis of the baryon oscillations and Type Ia 
supernovae observations \citep{SeoEisenstein:2003}.

%%%%%%%%%%%%%%%%%%%%%%%%%%%%%%%%%%%%%%%%%%%%%%%%%%%%%%%%%%%%%%%%%%%%%%%%%%%%%%%
%%%%%%%%%%%%%%%%%%%%%%%%%%%%%%%%%%%%%%%%%%%%%%%%%%%%%%%%%%%%%%%%%%%%%%%%%%%%%%%
\section{Conclusion}
\label{sec:summary}
%%%%%%%%%%%%%%%%%%%%%%%%%%%%%%%%%%%%%%%%%%%%%%%%%%%%%%%%%%%%%%%%%%%%%%%%%%%%%%%
%%%%%%%%%%%%%%%%%%%%%%%%%%%%%%%%%%%%%%%%%%%%%%%%%%%%%%%%%%%%%%%%%%%%%%%%%%%%%%%
We have studied the cross-correlation between CMB (the RS effect) and
large-scale structure traced by WL.
We have developed a simple analytical model based upon the time derivative of
non-linear matter power spectrum, and tested its validity analytically
with 3rd-order PT as well as numerically with $N$-body simulations.

We have shown that all-sky cosmic-variance limited CMB and deep
(shallow) WL surveys can yield a $50\sigma$ ($10\sigma$) detection of
the {\it non-linear} CMB-WL correlation for $l_{\rm max}=10^4$. 
The forthcoming surveys are not expected to yield significant detections.
We expect $\sim 1.5\sigma$ from Planck+LSST and $0.1\sigma$ from ACT+CFHTLS.

The change of the sign of $C_l$ at the cross-over, $l\sim 800$, offers
a unique probe of the nature of DE.
Sensitivity of the CMB-WL correlation to DE turns out to be not
very strong; however, as the direction of degeneracy on $w_0-w_1$ is
different from that of the baryon oscillations
and Type Ia supernovae, it may provide an independent
cross-check of the results from these observations. 

%%%%%%%%%%%%%%%%%%%%%%%%%%%%%%%%%%%%%%%%%%%%%%%%%%%%%%%%%%%%%%%%%%%%%%%%%%%%%%%
%%%%%%%%%%%%%%%%%%%%%%%%%%%%%%%%%%%%%%%%%%%%%%%%%%%%%%%%%%%%%%%%%%%%%%%%%%%%%%%
\section*{Acknowledgments}
%%%%%%%%%%%%%%%%%%%%%%%%%%%%%%%%%%%%%%%%%%%%%%%%%%%%%%%%%%%%%%%%%%%%%%%%%%%%%%%
%%%%%%%%%%%%%%%%%%%%%%%%%%%%%%%%%%%%%%%%%%%%%%%%%%%%%%%%%%%%%%%%%%%%%%%%%%%%%%%
We thank Olivier Dor{\'e}, Joe Hennawi, Ravi Sheth, and David Spergel
for useful discussions and comments on the paper.
The work is supported by Grant-in-Aid for Scientific Research on Priority Areas
No. 467 ``Probing the Dark Energy through an Extremely Wide \& Deep Survey 
with Subaru Telescope'' and by The Mitsubishi Foundation. The numerical 
computation is done by cluster computers in Nagoya University. EK acknowledges
support from an Alfred P. Sloan Fellowship.

\bibliography{bibdata} 
\bibliographystyle{apj}
\end{document}